# Measuring the Boltzmann constant by mid-infrared laser spectroscopy of ammonia

S Mejri[1,2,*], P L T Sow[2,1], O Kozlova[3], C Ayari[4], S K Tokunaga[1,2], C Chardonnet[2,1], S Briaudeau[3], B Darquié[2,1], F Rohart[5] and C Daussy[1,2,†]

[1] *Université Paris 13, Sorbonne Paris Cité, Laboratoire de Physique des Lasers, F-93430 Villetaneuse, France*

[2] *CNRS, UMR 7538, LPL, F-93430 Villetaneuse, France*

[3] *Laboratoire Commun de Métrologie LNE-CNAM, F-93210, La Plaine Saint-Denis, France*

[4] *Univ Tunis, Ecole Super Sci & Tech Tunis, Lab Dynam Mol & Mat Photon, Tunis 1008, Tunisia*

[5] *Laboratoire de Physique des Lasers, Atomes et Molécules, UMR CNRS 8523, Université de Lille, F-59655 Villeneuve d'Ascq cedex, France*

E-mail: christophe.daussy@univ-paris13.fr

**Abstract:**

We report on our on-going effort to measure the Boltzmann constant, $k_B$, using the Doppler broadening technique on ammonia. This paper presents some of the improvements made to the mid-infrared spectrometer including the use of a phase-stabilized quantum cascade laser, a lineshape analysis based on a refined physical model and an improved fitting program increasing the confidence in our estimates of the relevant molecular parameters, and a first evaluation of the saturation parameter and its impact on the measurement of $k_B$. A summary of the systematic effects contributing to the measurement is given and the optimal experimental conditions for mitigating those effects in order to reach a competitive measurement of $k_B$ at a part per million accuracy level are outlined.

---

[*] Present address:, LNE-SYRTE, Observatoire de Paris, CNRS, UPMC, 61 Avenue de l'Observatoire, 75014 Paris, France
[†] Author, to whom any correspondence should be addressed.

## 1. Introduction

Molecular spectroscopy plays an important role in precision measurement physics. It can be used to test fundamental symmetries [1-3], to determine absolute values of fundamental constants as well as their possible variation in time [4-6]. Here we report the latest progress on an ongoing experiment to spectroscopically measure $k_B$ the Boltzmann constant. A molecular (or atomic) absorption line of a gas at thermodynamic equilibrium exhibits a Doppler broadened profile which reflects the Maxwell-Boltzmann velocity distribution of gas particles [7]. In conjunction with some highly accurate modeling of this line profile, it is possible to use the data to retrieve the Doppler width $\Delta\nu_D$, whose relation to $k_B$ is given by [8-15]:

$$k_B = \frac{mc^2}{2T}\left(\frac{\Delta\nu_D}{\nu_0}\right)^2 \qquad (1)$$

where m is the molecular mass, $\nu_0$ the central frequency of the molecular line and $T$ the gas temperature. We have chosen to apply the "Doppler Broadening Thermometry" (DBT) on a cell of ammonia [16-19]. The probed transition of interest is the rovibrational $\nu_2$ saQ(6,3) line, found near 10 μm in the mid-infrared.

This paper begins by outlining some of the improvements made to the spectrometer, including the use of a phase-locked 10-Hz line width Quantum Cascade Laser (QCL) as the light source (see section 2). This is followed in section 3 by a discussion of our latest lineshape analysis. The purpose of this work was to refine the physical model used to fit the data and extract accurate line shape parameters by using an improved fitting code. Some old data is re-analyzed using various models that account for speed-dependent as well as Lamb-Dicke-Mössbauer (LDM) effects. We revisit the way we consider the underlying hyperfine structure of the probe transition and, for the first time, distortions of the lineshape due to the finite detection bandwidth of the spectrometer are also included in the analysis. In section 4, we then present a first evaluation of the saturation parameter in our experimental conditions and discuss its impact on the accuracy of the Doppler width measurement. By choosing the right parameter space, the contributions of the various systematic effects could be rendered negligible. Section 5 gives a revised uncertainty budget and presents the optimal experimental conditions for mitigating those effects in order to reach a competitive measurement of $k_\mathrm{B}$ at a part per million accuracy level.

**2. The 10 μm laser spectrometer**

The experimental setup is shown in figure 1. Although the final line half-width to be measured for the DBT is on the order of 50 MHz, it is essential for the studies of systematic effects that the spectrometer be able to record lines at much higher resolutions using saturated absorption spectroscopy. Three different cells are thus used, through which the light is sent in different configurations to be detailed below. Also to be discussed later are the corresponding examples of spectra, which are shown in figure 2.

The spectrometer uses the secondary frequency standard around 10 μm, a $CO_2$ laser stabilised on a saturated absorption line of $OsO_4$ as shown in figure 1(a) [20, 21]. The $CO_2$ laser frequency stabilization scheme is described in reference [21]: two sidebands are generated on each side of the $CO_2$ laser carrier with a tunable radio-frequency electro-optic modulator (RF-EOM). One sideband is tuned in resonance with an $OsO_4$ saturated absorption line detected in transmission of a 1.6-m long Fabry-Perot cavity on a liquid nitrogen-cooled HgCdTe detector (figure 1(a)). A double frequency modulation applied to the RF and in turn to the sideband allows the cavity to be frequency locked to the sideband and the sideband to the molecular line respectively. The laser spectral width measured by the beat note between two independent lasers is smaller than 10 Hz and the laser exhibits frequency instability of 0.1 Hz for a 100 s integration time.

Broad frequency tunability is achieved by coupling the $CO_2$ laser to a microwave electro-optic modulator (MW-EOM) which generates two sidebands (SB+ and SB-) tunable from 8 to 18 GHz on each side of the fixed $CO_2$ laser frequency. The intensity ratio between these two sidebands and the laser carrier is about $10^{-4}$. To increase laser power for spectroscopy a continuous wave near-room-temperature single-mode distributed feedback (DFB) quantum cascade laser (QCL) is coherently phase-locked to one of the two tunable sidebands (figure 1 (b)). Details of the locking characteristics are outlined in reference [22]. The phase lock loop both narrows the line width of the DFB-QCL and references its absolute frequency to the $OsO_4$ frequency grid [22]. The DFB laser is a 10.3 μm QCL (commercial device from Alpes Laser) typically operated at a temperature of 243K, a current of 1 A and exhibits a maximum output power of 60 mW. The beat signal between the QCL and one sideband of the $CO_2$ laser is detected on the detector of figure 1(b). The phase-error signal is generated by using a frequency mixer to compare the phases of the beat signal and of a reference (at typically 11 MHz). It is then processed by a proportional integrator servo-loop to obtain the correction to be applied to the QCL's current. The stability and accuracy of the $CO_2$ laser standard are transferred to the QCL resulting in a line width narrowed by about 4 orders of magnitude down to the 10-Hz level [22].

Figure 1(c) shows one of the two sub-Doppler parts of the apparatus, developed to record the hyperfine structure of the $NH_3$ saQ(6,3) rovibrational line, a source of systematic error when measuring $k_B$ using the DBT (see [18] for details). A 3 m-long Fabry–Perot cavity (FPC) is used to reach the 10-kHz resolution required for this study. The hyperfine components of the molecular line are recorded using a 1-kHz frequency modulation (2-kHz depth) applied on the $CO_2$ laser sideband via the MW-EOM. The resolution achieved is limited by the combination of gas pressure, transit time and modulation settings.

Figure 1(d) shows the other sub-Doppler part, developed for determining the saturation intensity of the probed line. For the purpose of studying saturation effects, to be outlined later, we specifically do not wish to resolve the hyperfine structure. Thus, this setup is built to yield a lower resolution than that in part (c). The saturated absorption signal is detected in transmission of a ~0.4 m-long absorption cell. The beam is retroreflected in a double pass configuration, and amplitude modulation of up to 2 kHz is applied using a chopper. The laser power is around 0.3 mW inside the absorption cell, and typical gas pressures range from 1 to 10 Pa.

Figure 1(e) represents the apparatus used to measure the Doppler broadened profile for determining $k_B$ [17]. The setup allows either laser (the $CO_2$ or the DFB-QCL) to be used for the measurement, as required. When using the $CO_2$ laser, the output of the MW-EOM is sent through a 3 cm-long FPC in order to filter out the carrier and the unwanted sideband SB+ as well as to stabilize the intensity of the transmitted red detuned sideband SB−. Spectroscopy is performed in transmission of an absorption cell whose length can be interchanged between 37 cm in a single-pass configuration (SPC) and 3.5 m in a multi-pass configuration (MPC). For an accurate control of the temperature, this cell is placed inside a copper thermal shield which is itself inside an enclosure immersed in a large thermostat ($1 \times 0.8 \times 0.8$ m$^3$) filled with an ice-water mixture, which sets the temperature very close to 273.15 K (see [17] for details). The gas temperature is measured with three small 25 Ω Hart capsule Standard Platinum Resistance Thermometers in thermal contact with the absorption cell. These thermometers are calibrated at the triple point of water and at the gallium melting point at the Laboratoire Commun de Métrologie LNE-CNAM. In order to measure the cell temperature, they are monitored with a resistance measuring bridge continuously calibrated against a resistance standard with a very low-temperature dependence. The probe beam is amplitude-modulated at $f = 40$ kHz (via either the MW-EOM or an acousto-optic modulator (AOM, see figure 1(d)) depending on the beam used) for noise filtering and signal is obtained after demodulation at f. The laser frequency is tuned close to the desired saQ(6,3) molecular resonance and scanned (typically over 500 MHz) to record the Doppler profile.

Our first spectra recorded with the DFB-QCL at ammonia pressures up to 149 Pa in the SPC are displayed in figure 2(a). They are about 100 MHz wide, recorded over 500 MHz in steps of 500 kHz with a 30 ms integration time (lock-in amplifier time constant) per point. Compared to our previous spectrometer based solely on a $CO_2$ laser only, the time needed to record absorption spectra (previously limited by the complexity of the $CO_2$ laser frequency tuning technique) is expected to be greatly reduced, leading potentially to a reduction in the statistical (type A) uncertainty on $\Delta\nu_D$. Furthermore, the increased intensity allows power-related systematic effects to be studied for the first time. In particular, it allows the first evaluation of the saturation parameter in our experiment and its impact on the determination of $k_B$, as detailed in section 4.

**3. The line shape modeling**

In order to extract the Doppler component of the width, the measured line shape is fitted to a physical model. The choice of model is critical. When implemented in the fitting routine, an inappropriate model will significantly shift the measured value of $k_B$, and is thus a clear source of potential systematic errors. Additionally, the choice of model is not obvious, as various physical effects, present in varying degrees depending on the experimental setup and conditions, rapidly modify the true observed profile [13, 23-26]. This section describes our latest progress with regards to modeling and fitting the line shape. The most significant improvement was in fact the use of a more sophisticated fitting program. This new fitting code is based on a Fourier transform technique, which allows for a larger numerical robustness as the time domain line shape has an exact expression [13, 27, 28]. Its use led to a major reduction of the computation time, allowing us to fit scans individually (instead of having to average them first). We were therefore able to re-analyze old data (presented in Ref. [29], 482 spectra for pressures ranging from 1.7 up to 20 Pa in a 500 MHz spectral range) to refine the line shape model and to assess the validity of the associated molecular parameters. This section describes, in turn, the new fitting code including the use of a new line profile, the inclusion of distortions to this line shape due to the detection bandwidth and the use of a sum of 5 lines to mimic the underlying hyperfine structure. Compared to typical pressure conditions (~1 Pa) for the measurement of $k_B$, the analyzed spectra were taken at rather large pressures in order to enhance collisional effects. The point here is to measure accurate line shape parameters to be used then when analyzing data recorded to measure $k_B$. For all numerical adjustments carried out throughout this section, the Doppler broadening is thus fixed, but to different values within a 100 ppm range around the value determined by the temperature of the gas. The uncertainty of each line shape parameter (reported in table 1) will thus take into account an overestimated uncertainty of 100 ppm for $k_B$.

We recall that in previous work we showed that our experiment operates in conditions where the contributions to the width of our observed line shape in descending order of magnitude were: $\Delta\nu_D$, homogeneous pressure broadening, speed-dependent distortions, and finally LDM narrowing. So far, we had therefore neglected the final contribution and used the speed-dependent Voigt profile (SDVP) [29, 30]. In this work, we report on the inclusion of this final effect into our model. The LDM effects result in a reduction of the Doppler width due to velocity changing collisions and can be modeled, assuming soft collisions between molecules, by the Galatry profile (GP). When combined with the speed-dependent distortions of the Lorentzian, the line shape model is usually called the speed dependent Galatry profile (SDGP) [31].

As before, the more significant speed dependent effects have been considered here via the so-called hypergeometric model of Berman and Pickett [31-33], for which collisional broadening and frequency shift are proportional to some power of the relative speed $v_r$ in a molecule-molecule collision and given by $\Gamma \sim (v_r)^m$ and $\Delta \sim (v_r)^n$, respectively. Speed dependent effects lead to a line narrowing that can be easily and reliably analyzed given a sufficient signal to noise ratio. However, the analysis of the speed dependence on the frequency shift is more delicate, namely because the resulting line asymmetry is such that the line center and absorption peak no longer coincide. In the case of the saQ(6,3) $NH_3$ line, this analysis is as difficult as the frequency shift is very weak (about 1-2% the broadening value).

The line shape analysis concerning the speed dependence was conducted as follows[‡]. The line shape Fourier transform is a time domain signal, proportional the dipole correlation function $\Phi(\tau)$ and given by [31-35] :

$$\Phi(\tau) = \int \exp\{i\omega_0\tau + i\Delta(v_a)\tau - \Gamma(v_a)\tau\}\text{sinc}(kv_a\tau)f(v_a)\,dv_a, \quad (2)$$

with, for self-collision case considered here,

$$\Gamma(v_a) = \Gamma_0 2^{-m/2} M(-m/2; 3/2; -(v_a/v_{a0})^2) \text{ and } \Gamma_0 = \langle\Gamma(v_a)\rangle_{v_a}, \quad (3)$$

and $\quad \Delta(v_a) = \Delta_0 2^{-n/2} M(-n/2; 3/2; -(v_a/v_{a0})^2) \text{ and } \Delta_0 = \langle\Delta(v_a)\rangle_{v_a}. \quad (4)$

---

[‡] In order to simplify the discussion, LDM contributions [13, 27], actually included in the fitting code, are omitted in following equations.

Here, $\omega_0$ is the unperturbed line frequency, $k = \omega_0/c$ the wavenumber, $\text{sinc}(x) = \sin(x)/x$, $f(v_a)$ the Maxwell-Boltzmann distribution of absorber speeds $v_a$, $v_{a0}$ the most probable absorber speed, $\Gamma_0$ and $\Delta_0$ the mean broadening and frequency shift and $M(a; b; x)$ the confluent hypergeometric function. A reliable determination of $\Delta_0$ is, in general, quite difficult to do because it is tiny. The determination of the speed exponent $n$ is equally challenging, as it is very sensitive to the base line. A strategy consists in re-writting the signal as:

$$\Phi(\tau) = \int \exp\{i[\omega_0 + \Delta_0^a]\tau + i[2^{-n/2}M(-n/2; 3/2; -(v_a/v_{a0})^2) - 1]\Delta_0^b \tau$$
$$- 2^{-m/2}M(-m/2; 3/2; -(v_a/v_{a0})^2)\Gamma_0 \tau\} \text{sinc}(kv_a\tau)f(v_a)\,dv_a$$

(5)

where first and second imaginary terms refer to line frequency shift and asymmetry, respectively. $\Delta_0^a$ and $\Delta_0^b$ are strongly correlated with the speed exponent $n$ but their pressure dependences $\delta_0^a = d\Delta_0^a/dP$ and $\delta_0^b = d\Delta_0^b/dP$ must be equal. So, the retained $n$-value must be such that $\delta_0^a = \delta_0^b = \delta_0$ after analyzing data taken at various pressures. For this purpose, several attempts involving different fixed $n$-values were done. In the fitting procedure the Doppler broadening was fixed and adjusted parameters were: line amplitude $A_0$, line frequency ($\omega_0 + \Delta_0^a$), broadening $\Gamma_0$ and speed exponent $m$, frequency shift $\Delta_0^b$ as well as a linear base line.

Moving on to the effect of velocity changing collisions [31, 36, 37], LDM effect is nearly negligible. Indeed, the mass diffusion coefficient measured for $^{14}NH_3$ in a classical transport study is $D_{NH_3} = 1.58(3) \times 10^{-5} m^2 s^{-1}$ at 1 atm and 273.15 K [29] leading to a frequency of velocity-changing collisions of $\beta_{kin} = 13.2(3)$ kHz/Pa. As was done for other molecular systems [35, 38-40], a conservative upper limit of the frequency of velocity-changing collision to be used in the line shape has therefore been set to 0.10 $β_{kin}$. It is very small, in particular compared to the ~120 kHz/Pa (see below) collisional broadening. In the case of NH$_3$ self-collisions, the influence of velocity changes on the line shape is thus completely smeared out by the large dipole relaxation rate [31, 35, 37, 38]. Moreover, at that level, no difference is expected between the Galatry and Rautian models [38, 41]. The former is more suitable for Fourier transform based numerical code [35] and was thus used in this study.

In addition to these effects, the model also comprises terms to account for technical distortions. We have recently shown that the finite bandwidth of any detection system (in our experiment limited by the lock-in detection bandwidth) can cause large distortions to the measured line shapes [42]. Simulations showed that the contribution of this systematic effect impacts the Doppler width at a level larger than 10 ppm in our typical experimental conditions. In [42], we have developed a model to account for this effect. We include it here in the absorption line shape. Note however that we approximate our spectrometer frequency sweep consisting of a series of discrete steps by a continuous sweep of the laser frequency.

Finally, we revisit the way we treat the hyperfine structure. The saQ(6,3) line has been chosen because it is a well-isolated rovibrational line. However, owing to the non-zero spin values of the N and H nuclei, hyperfine structure is present in the Doppler profile. Saturated absorption spectroscopy (see [18, 43] and spectrum on figure 2(c) recorded with the apparatus shown on figure 1(c)) provided an accurate knowledge of the hyperfine structure of this rovibrational line, which can be included in the absorption line shape model. In our previous studies, distortions due to underlying hyperfine structure were added as a correction, following a study showing that the effect was in fact a systematic overestimation of the Doppler width [29, 30]. A more realistic model of the saQ(6,3) $^{14}$NH$_3$ line has been considered in the new code: the main line is surrounded by 4 weaker additional lines (± 0.12 and ± 0.57 MHz away with respective amplitudes 4.2% and 2.5% of that of the main line, as deduced from [18]), amplitudes, frequencies and relaxation parameters being constrained to corresponding parameters of the main line. Although no major changes were noted following this enhancement (see below), it is reassuring to find compatible results in view of the many other changes made to the data analysis since the original correction factor was calculated.

Values determined with Voigt, Speed Dependent Voigt, Galatry and Speed Dependent Galatry profiles using the new code are reported in Table 1 along with those determined previously in [29] for the Voigt, Speed Dependent Voigt and Galatry profile. These results were obtained from 482 individual spectra recorded in a 500 MHz span at pressures ranging from 1.7 up to 20 Pa, with signal to noise ratios from 1000 to more than 3200. The aim of our project being an independent determination of $k_B$, as mentioned above, the uncertainty of each line shape parameter takes into account an overestimated uncertainty of 100 ppm for $k_B$ (that given by the CODATA is only 0.9 ppm) corresponding to an uncertainty of 50 ppm for the fixed value of the Doppler broadening. Incidentally, those uncertainties are dominated by the limited signal-to-noise ratio, not by the 100 ppm $k_B$ uncertainty. A typical spectrum recorded at 19.4 Pa is displayed in Fig. 3 together with residuals obtained after fitting data with a VP and a SDGP. In the fitting procedure, each individual point in a spectrum has a weight proportional to the inverse of the local transmittance, to account for amplitude noise.

For all profiles, the central line frequency is in agreement with the value already obtained by saturated absorption techniques 28 953 693.9(1) MHz [18]. As expected, the SDGP collisional broadening is larger than the Voigt one. Indeed, for a given measured width, fitting a broadening as well as a competing narrowing term would make the code converge to a larger value for the broadening parameter compared to fitting a broadening mechanism alone. This difference is on the order of 3%, because absorber and collisional partners have the same mass and the speed exponent $m$ (0.35) is small. By comparison with [29], more important differences are observed for the speed dependent line frequency shift, namely for the exponent *n*, (−5.0(5) and −3.8(3), respectively (see Table 1)). We attribute this difference to the improved robustness of the code. In particular, we are now able to fit scans individually, instead of having to average them first. This means scans with different sweep directions are fit independently, which is important when considering distortions due to detection bandwidth. Indeed those result in a +/- 0.25 MHz overall shift of the profile, the sign depending on the sweep direction. Fitting averaged data ends up affecting the extracted value for *n*, accounting for the observed difference in the results of the two analyses.

In both cases, the determination of *n* compared to other parameters remains imprecise (~10%). We recall that this parameter is responsible of a line asymmetry that, in the present case, is tiny because the frequency shift parameter $\delta_0$-value is only about 1% of the broadening parameter $\gamma_0$-value (see Table 1). $\delta_0$-values obtained from pure VP and SDGP (1.71(15) and 1.11(20) kHz/Pa, respectively) differ by a factor 1.5. Indeed as already quoted, the asymmetry is such that the line center and the absorption pick no longer coincide. Finally, we note that the analysis using the Galatry profile both in this work and in [29] is not valid as it ignores speed dependent effects. This is demonstrated by the frequency of velocity-changing collisions $\beta_G$, which was found to be ~25 kHz/Pa , 2 times larger than the kinetic diffusion parameter ($\beta_{kin} = 13.2(3)$ kHz/Pa), despite the latter being the upper limit of the former [31, 35, 37, 38].

## 4. Saturation of the molecular transition and the impact on determining $k_B$

Saturation of an absorption line leads to a deviation of the line shape from the Beer-Lambert law, potentially causing a systematic effect on the $k_B$ measurement. We estimate the saturation parameter of the saQ(6,3) rovibrational transition for our experimental conditions. One direct method of measuring the saturation parameter is by saturated absorption. The expression of the absorption coefficient of one of the two counterpropagating beams (beam +) of a saturated absorption setup is given in the weak field limit (valid under our experimental conditions) by [44] :

$$\alpha_s^+(\omega) = \frac{n_{ab}^0}{\left(\frac{c\varepsilon_0(E_0^+)^2}{2\hbar\omega}\right)} \frac{1}{\sqrt{\pi}\Delta\nu_D} e^{\left[-\left(\frac{\omega-\omega_0}{\Delta\omega_D}\right)^2\right]} \left\langle (\Omega^+)^2 \left[1 - (\Omega^+)^2 2T_1 T_2 - (\Omega^-)^2 2T_1 \operatorname{Re}\left[\frac{1}{T_2 - i(\omega-\omega_0)}\right]\right]\right\rangle \quad (6)$$

where $n_{ab}^0$ is the total population difference per unit volume between the two involved levels, $\omega_0$ is the resonant angular frequency, $E_0^\pm$ are the electric field amplitudes of the two counterpropagating waves and $\Omega^\pm = \frac{\mu E_0^\pm}{2\hbar}$ are their Rabi frequencies where $\mu = \langle a|\vec{\mu}|b\rangle$ is the electric dipole matrix element. $T_2 = \frac{1}{\gamma_{ba}}$ where $\gamma_{ba}$ is the optical dipole relaxation rate and $T_1 = \frac{1}{2}\left(\frac{1}{\gamma_a} + \frac{1}{\gamma_b}\right)$ where $\gamma_{a,b}$ are the population decay rates. For degenerate levels, the products of Rabi frequencies have to be summed over all possible quantum sub-states leading to :

$$\alpha_s^+(\omega) = \alpha(\omega)\left[1 - \frac{I^+}{2I_s} - \frac{I^-}{2I_s}f(\omega)\right] \quad (7)$$

where $\alpha(\omega)$ is the absorption coefficient in the absence of saturation effect, $I^\pm = \frac{c\varepsilon_0}{2}\left|E_0^\pm\right|^2$ are the laser intensities of the two counterpropagating waves, f the homogeneous saturated absorption profile and $I_s$ the saturation intensity given by [44] :

$$I_s = \frac{8\pi h\nu^3}{c^2} \cdot \frac{1}{A_{ba} \cdot g_b} \cdot \left[\frac{1}{9A_i^X g_\pm}\right] \frac{1}{4T_1 T_2} \quad (8)$$

with $A_i^X$ an angular factor calculated in [45], $A_{ba}$ the Einstein coefficient, $g_b$ the degeneracy factor of the upper level and $g_\pm$ a geometrical factor equal to ½ for a Gaussian laser beam (assumed here with a ~5 mm waist) and to 1 in the limit of a homogeneous power over the whole transverse laser beam profile [44]. Under typical experimental conditions (~1 Pa), $\frac{1}{2\pi T_2} = 120$ kHz (see Table 1) and $T_1$ is estimated to be 25% of $T_2$ [46, 29].

From eq. (8) and using the Einstein coefficient found in the HITRAN database [47] the saturation power for the saQ(6,3) rovibrational line is estimated at 10 mW.

This estimate is compared to a measured value extracted from the ratio of linear to saturated absorption amplitudes, hereafter referred to as the 'contrast' $\mathcal{C}$. From eq. (7), we find that $\mathcal{C}$ is given by $I_0/2I_s$ assuming $I^+ \approx I^- \approx I_0$ (corresponding to the contrast that would be obtained in the case of an optically thin medium with counterpropagating beams of intensity $I_0$), giving a simple method of estimating $I_s$. The measurements are made using parts (a), (b) and (d) of the setup as shown in figure 1. The molecular vapor is probed with a Gaussian beam of ~5 mm waist and powers up to 0.3 mW. Saturated absorption spectra have been recorded in the 0.4 m-long absorption cell (figure 1 part (d)). Note that the characteristic widths of the linear and saturated absorption signals are very different. Measurements using frequency modulation therefore become complicated because amplitudes are different functions of the modulation depth. We therefore revert to amplitude modulation despite the reduced signal to noise ratio. A typical saturated absorption spectrum is shown in figure 2(b) for a power at the entrance of the absorption cell of $P_0 \sim 0.3\ mW$ and a gas pressure of $\sim 1\ Pa$. We find $\mathcal{C}$ to be about 2 to 3%, implying a saturation power of about 5 to 8 mW, compatible with that calculated from eq. (5). Note that the typical absorptions at 1 Pa after retroreflection being of the order of 20%, we are not quite in the optically thin medium case, but our estimate provides the right order of magnitude.

We recall that $I_s$ was estimated in order to set an upper bound on a potential systematic effect when measuring $k_B$. Indeed, if our laser power is anywhere near $I_s$ during the measurement, it could potentially deform the line shape causing a systematic error. We now estimate this systematic error by simulating saturated Doppler spectra, which we then fit to unsaturated profiles. The absorption coefficient for linear absorption in the presence of saturation is given by [48]:

$$\alpha(\omega) = -\frac{\sqrt{1+\frac{I(\omega,z)}{I_s}}}{I(\omega,z)} \frac{dI(\omega,z)}{dz} \quad , (9)$$

which after integration over the absorption length $L$ leads to :

$$\left[ \text{Ln}\left( \frac{\sqrt{1+\frac{I(\omega,z)}{I_s}}-1}{\sqrt{1+\frac{I(\omega,z)}{I_s}}+1} \right) + 2\sqrt{1+\frac{I(\omega,z)}{I_s}} \right]_{I_0}^{I(\omega)} = -\alpha(\omega).L \quad (10)$$

Under weak saturation (valid in our measurement conditions), a first-order expansion in $\frac{I_0}{I_s}$ leads to the following generalization of the Lambert–Beer law :

$$I(\omega) \simeq I_0 \left[ 1 - \frac{I_0}{2I_s}\left(e^{-\alpha(\omega)L} - 1\right) \right] e^{-\alpha(\omega)L} \quad (11)$$

which can also be obtained directly from (7) by eliminating the terms associated with the return beam. We now use this expression to generate 'saturated' Beer-Law-Voigt profiles (with $\Delta\nu_D = 50\ MHz$ and $\Gamma = 120$ kHz/Pa see section 3). Fitting those to the unsaturated equivalent profile (in the same way we treat our experimental data) results in a significant underestimation of the Doppler linewidth. Under typical experimental conditions (pressure of ~ 1 Pa and power of ~1 µW [18]) we estimate this systematic effect to be about ~ -25 to -50 ppm.

To overcome this, future measurements will have to be recorded at a lower saturation parameter. In order to reduce this systematic effect below 1 ppm, saturation parameters as low as $4.10^{-6}$ should be used, corresponding to laser power below 0.02 µW when working at pressures close to 1 Pa. Such low laser power will not be an issue for the signal-to-noise ratio as the photodetector noise remains below the laser amplitude noise level even at a laser power as small as of 0.02 µW.

**5. Conclusions and prospects**

Table 2 summarises the uncertainties associated with measuring $k_B$ using the DBT. It is a revision of the total uncertainty budget, so some of the contributions are not detailed in this work. Their description can be found in [5]. The combined type B uncertainty on the thermometers' measurement is 0.3 mK. Temperature inhomogeneity across the cell and temperature drifts on the timescale of a single spectrum yield an additional uncertainty of 0.23 mK.

For the Doppler width determination, no systematic effect due to the laser beam geometry, the natural width of rovibrational levels, gas composition, linearity and accuracy of the laser frequency scale or laser linewidth is expected. Attempts to observe systematic effects due to detection chain nonlinearity (including photodetector and locking amplifier) were unsuccessful at a 10ppm level [18]. The laser beam modulation and the unresolved hyperfine structure are responsible for an overestimation of the Doppler width that can be corrected for at the level of 0.02 ppm and 0.013 ppm, respectively.

As reported in section 3, the finite bandwidth of a detection system causes a distortion of the measured line shape. One could include this effect in the fitting routine of the final $k_B$ measurement, as was done in section 3. Although this approach is suitable for analyzing the line shape and extracting its parameters ($m$, $n$ and $\beta$) at a few-percent level at high pressures, it is impractical for measuring $k_B$, as it would require accurate knowledge of the lock-in detection $2^{nd}$ order filter time constant $\tau_D$ and the quality factor $Q$. Instead, we propose to operate in a region of parameter space where this effect is below 1 ppm, which we now know can be attained using a Butterworth second-order filter (−12 dB/octave roll-off and $Q = \sqrt{1/2}$) as detailed in [42].

As reported in section 4, saturation also causes a line shape distortion and is thus a potential systematic effect when determining $k_B$. As with the detection bandwidth distortions, the solution here is again to move to a parameter space where the effect is negligible. For this, the saturation parameter will have to remain below $4.10^{-6}$, which means the laser power entering the cell must be kept below 0.02 µW at 1 Pa.

Having thus eliminated essentially technical effects, the main remaining contributions to the error budget are the parameters *m* and *n*. As reported in ref [29], following the line-absorbance based analysis method proposed by Castrillo et al. [49], we have estimated this contribution to be of 1.8 ppm, provided frequency scans span at least 500 MHz and pressures are between 1 and 2 Pa, the latter to ensure high enough signal-to-noise ratios.

**Acknowledgements**

This work is funded by CNRS, Agence Nationale de la Recherche (ANR QUIGARDE 2012 ANR-12-ASTR-0028-03, ANR-10-LABX-005 through the Programme d'Investissement d'Avenir), Labex First-TF (ANR 10 LABX 48 01) and AS GRAM. PhLAM laboratory is also supported by Région Nord–Pas de Calais, Ministère de l'Enseignement Supérieur et de la Recherche and Fond Européen de Développement Economique des Régions (FEDER). The authors would like to thank C. Martin, F. Sparasci and V. Vidal from Laboratoire Commun de Métrologie LNE-Cnam for platinum resistance thermometers calibrations and assistance in temperature data acquisition. The authors also thank Ch. J. Bordé for many fruitful discussions over the course of this project.

Figure 1: Experimental setup based on (a) a CO2 laser and (b) a Quantum Cascade Laser (QCL) for (c) (respectively (d)) saturated absorption spectroscopy in a Fabry–Perot cavity (FPC) (respectively in a cell) and (e) linear absorption spectroscopy (MW-EOM, microwave electro-optic modulator; AOM, acousto-optic modulator; SB, sideband; Detect., detector).

Figure 2 : Spectra of the $\nu_2$ saQ(6,3) line of $NH_3$ recorded (a) with the QCL, in linear absorption for pressures ranging from 5.3 Pa to 149 Pa over a span of 500 MHz (in a cell and in the single-pass configuration), (b) with the QCL, in saturated absorption over a span of 10 MHz and using an amplitude modulation (in a cell), (c) with the $CO_2$ laser, in saturated absorption over a span of 200 kHz and using a frequency modulation (in a Fabry–Perot cavity).

Figure 3 : $\nu_2$ saQ(6,3) line of $NH_3$ recorded at 19.4 Pa (signal to noise ratio of about 3 200) and normalized residuals (magnified by a factor 50) for weighted nonlinear least-squares fits with a VP and a SDGP. The frequency scale is offset by 28 953 690 MHz.

Table 1 : Line shape spectroscopic parameters (and their standard uncertainties) derived from various line profiles for the self-broadened $\nu_2$ saQ(6,3) line of $NH_3$ around 273.15 K, (a) this work , (b) ref. [29], (c) for a cell length L $\cong$ 37 cm, (d) frequencies shifted by 28 953 690 MHz.

Table 2 : Type-B uncertainty budget (k = 1) for the determination of the Boltzmann constant by the DBT at LPL for spectra recorded at a pressure of $\sim 1 Pa$ in the multi-pass configuration.

Figure 1

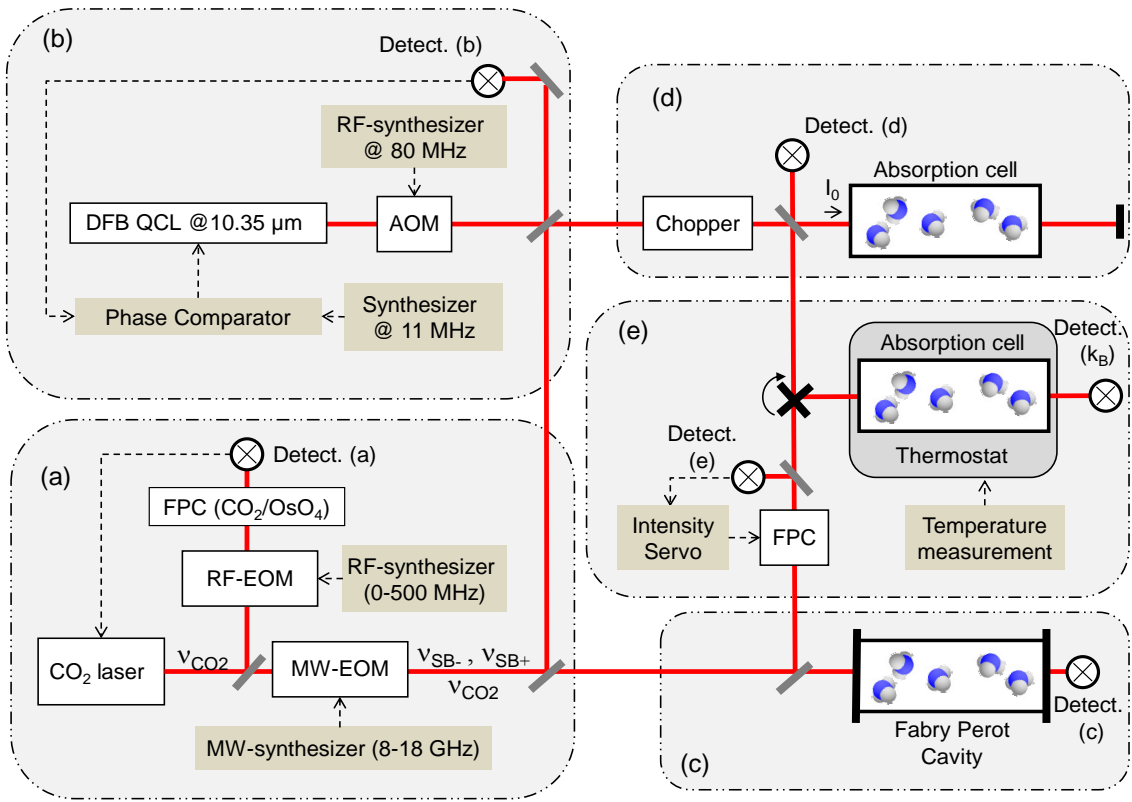

Figure 2

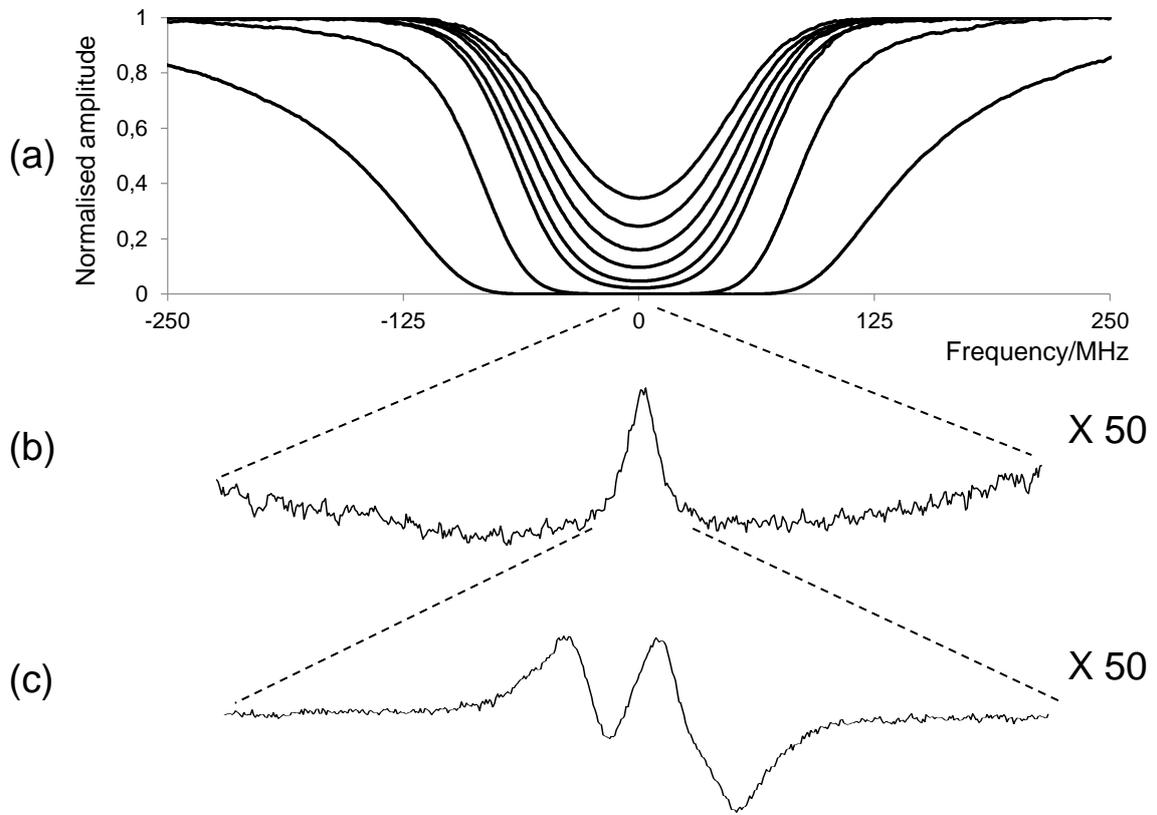

(a)

(b) X 50

(c) X 50

Figure 3

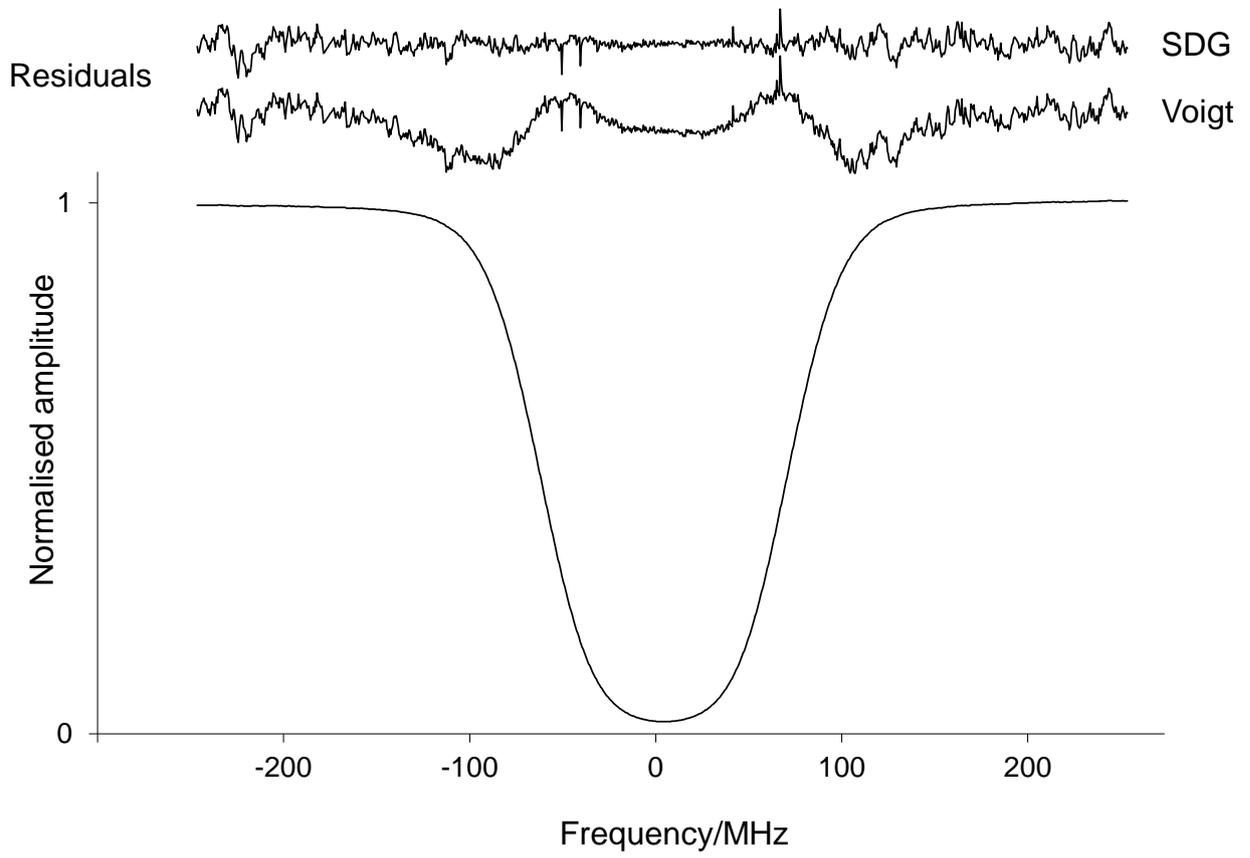

Table 1

| Parameter | Unit | Voigt profile | Galatry profile | Speed Dependent Voigt profile | Speed Dependent Galatry profile |
|---|---|---|---|---|---|
| | | | **This work** | | |
| $A_0$ | MHz/Pa | 18.143(3) | 18.209(4) | 18.205(4) | 18.208(4) |
| $\nu_0$ | MHz | 3.924(2) | 3.925(2) | 3.918(3) | 3.918(3) |
| $\delta_0$ | kHz/Pa | +1.7(2) | +1.7(2) | +1.1(2) | +1.1(2) |
| $\gamma_0$ | kHz/Pa | 117.9(3) | 122.4(4) | 121.5(3) | 121.7(3) |
| $\beta_G$ | kHz/Pa | 0 | 23.8(4) | 0 | $\equiv 1.32$ |
| $m$ | - | 0 | | +0.350(7) | +0.350(7) |
| $n$ | - | 0 | | -5.0(5) | -5.0(5) |
| | | | **From reference [29]** | | |
| $\delta_0$ | kHz/Pa | +0.99(4) | +1.01(4) | +1.2(1) | |
| $\gamma_0$ | kHz/Pa | 106 to 130 | 120.9(3) | 120(3) | |
| $\beta_G$ | kHz/Pa | 0 | 14 to 32 | 0 | |
| $m$ | - | 0 | 0 | +0.360(7) | |
| $n$ | - | 0 | 0 | -3.8(3) | |

Table 2

| Component | Relative uncertainty on $k_B$ (parts in $10^6$) | Comment |
|---|---|---|
| | **Temperature** | |
| Drift | 0.0007 | During 1 spectrum acquisition (1 min) |
| Inhomogeneity | 0.84 | Absorption cell inhomogeneity |
| $T$ measurement | 1.1 | LCM-LNE-CNAM estimation |
| | **Doppler width** | |
| Natural width | Negligeable | |
| Gas composition | Negligeable | |
| Laser beam geometry | No impact | In the absence of saturation effect |
| Laser linewidth | Negligeable | Deduced from the laser spectral width |
| Linearity and accuracy of the laser frequency scale | Negligeable | Deduced from the frequency tuning technique |
| Laser beam modulation | 0.04 | Numerical simulations |
| Hyperfine Structure | 0.03 | Deduced from the knowledge of the Hyperfine structure |
| Frequency sweeping rate | < 1 | With a Butterworth second-order filter |
| Saturation of linear absorption | < 1 | For laser power less than 0.02 µW |
| Self-broadened absorption line shape | 1.8 | Uncertainty on $m$ and $n$ |
| **Combined standard uncertainty** | **2.3** | **Root sum of squares** |